# SPEECH-BASED DEPRESSION PREDICTION USING ENCODER-WEIGHT-ONLY TRANSFER LEARNING AND A LARGE CORPUS

*Amir Harati, Elizabeth Shriberg, Tomasz Rutowski, Piotr Chlebek, Yang Lu, Ricardo Oliveira*

Ellipsis Health, San Francisco, CA, USA

{amir,liz,tomek,piotr,yang,ricardo}@ellipsishealth.com

## ABSTRACT

Speech-based algorithms have gained interest for the management of behavioral health conditions such as depression. We explore a speech-based transfer learning approach that uses a lightweight encoder and that transfers only the encoder weights, enabling a simplified run-time model. Our study uses a large data set containing roughly two orders of magnitude more speakers and sessions than used in prior work. The large data set enables reliable estimation of improvement from transfer learning. Results for the prediction of PHQ-8 labels show up to 27% relative performance gains for binary classification; these gains are statistically significant with a p-value close to zero. Improvements were also found for regression. Additionally, the gain from transfer learning does not appear to require strong source task performance. Results suggest that this approach is flexible and offers promise for efficient implementation.

***Index Terms***— transfer learning, encoder/decoder, depression, behavioral health, mental health

## 1. INTRODUCTION

Depression is a debilitating, prevalent condition that is often under-diagnosed [1][2]. Recent events such as COVID-19 [3] have increased the need for an automatic solution for the management of behavioral health disorders exacerbated by isolation and additional stressors [4]. Digital health solutions can play a role by improving the capacity for screening and monitoring for depression and other behavioral health conditions.

Within the field of digital health, spoken language technology offers distinct advantages. Speaking is natural and engaging for patients and requires only a microphone. Studies show that speech contains acoustic and language cues that can be captured by machine learning models to predict a speaker's behavioral health state, e.g. [5][6][7][8][9][10][11]. Common evaluations on the task of depression prediction have led to a growing number of studies focused on machine learning approaches [8][12][13][14][15].

Historically, feature engineering was an early and dominant approach for predicting depression in speech signals [9][16]. Sample features include voice quality [17][16], articulation [18][19][20], speech rate [19], and spectral [9] features. Advances in deep learning [21] have led to improved results in a range of affective and behavioral health tasks [22][23][24][25][26][27][28]. In deep learning the focus is to learn feature representation from data.

There has been increased research on the application of deep learning methods for the task of depression prediction. For example, in [10] and [29], the authors investigated different deep convolutional neural networks (CNN). In [8], a pretrained CNN (pretrained over an image classification task) was used as a feature extractor. In [12], the authors used a hierarchical model with transfer learning to achieve very good results on the AVEC [8] development set, although performance degraded when evaluated on the test set.

In this paper, we introduce a depression prediction approach that harnesses transfer learning but is also developed with scalability, low latency, and a more lightweight run-time system in mind. In order to achieve these goals, we explore a transfer learning approach that uses a lightweight encoder and that transfers only the encoder weights. The model is trained and evaluated on an appropriately large corpus for evaluating model robustness and discerning the performance gain from transfer learning. Evidence of robustness comes from test data performance that shows no difference from performance on development data (all partitions, including training data, use unique speakers).

Our dataset includes nearly 11,000 unique speakers. As a comparison, the AVEC/DAIC set [8] contains fewer than 300 total speakers. The size of our dataset is roughly two orders of magnitude larger in both speaker and session statistics than data used in major benchmarks [8][13][14][15].

Our speech samples come from human-computer interactions with a speech application. Gold standard labels for the machine predictions are based on a standard self-report instrument that is completed by each participant within each interaction session. Using this setup, we seek answers to two questions critical for real-world applications. First, we ask to what degree transfer learning can improve depression prediction performance over an end-to-end acoustic baseline model. Second, we explore whether transfer learning can be based on a weakly-performing source task. If so, the pretraining process can be simplified and development time can be significantly reduced.

## 2. METHOD

### 2.1. Data

We used a corpus of American English conversational speech collected by Ellipsis Health. We used this data set because we were unable to find a large enough publicly available depression-labeled speech corpus to allow suitable examination of our research questions. Due to privacy issues associated with health-related data, our data is currently proprietary. While we are in parallel investigating whether some data can be shared, the fact is that in the domain of mental health, there are currently no large, labeled sets of audio recordings publicly available for direct community comparisons. As noted earlier, these sets contain just a few hundred speakers to be used across all partitions.

Our goal is not to provide an absolute results benchmark but rather to show *relative* gains from a good baseline using our approach to transfer learning in this domain space. We believe that this provides value to the community interested in real-world applications. Note that even if we could access common data sets, comparison with results trained using a small dataset would not be fair.

Our corpus contains 10,932 unique speakers, 59% of whom are female and 41% of whom are male. Users ranged in age from 18 to 64 with an average age of 30. Questions were designed to elicit responses relating to the user's personal life, e.g. current concerns, home life, etc. Users spoke freely in response to questions within a session. On average, each session contains 354 seconds of audio.

During each session, the user also completed a Patient Health Questionnaire depression scale (PHQ-8) survey, a standard self-report instrument used for depression screening [30]. The PHQ-8 distribution was skewed; the majority of patients had a PHQ-8 score under 5, while only a small minority had a PHQ-8 score above 15. Within each session, the PHQ-8 was administered after the collection of speech samples. For binary classification task, following [30], we mapped the PHQ-8 score to positive for depression (+dep) if the output was 10 or above, and to negative for depression (-dep) if the output was below 10. Table 1 shows corpus statistics. The train, development, and test splits contain no overlapping speakers. The test and development splits are of approximately the same size and contain unique speakers; 43% of sessions in the train split are from recurrent users.

### 2.2. Acoustic Models

The acoustic model uses a neural network structure based on an encoder/decoder architecture [31]. We then use this baseline model to test the degree to which transfer learning can improve results. We also examine the impact of the accuracy of the transfer learning source task on the benefit from transfer learning. We study both binary classification and regression tasks. In all experiments, models were trained only on data from the train split. We stopped training once there was no more improvement on the development split. We then used the single outcome of that model on the test set for our evaluation.

**Table 1** Corpus statistics by partition and condition class. Train, development, and test data contain no overlapping speakers.

|  | Train | Train +dep | Test+Dev | Test+Dev +dep |
|---|---|---|---|---|
| Responses | 57835 | 16277 | 14534 | 3139 |
| Sessions | 12872 | 3606 | 3078 | 653 |

#### 2.2.1. End-to-End model

Our 'baseline' approach feeds features extracted from audio signals into a neural network. The end-to-end model is based on an encoder/decoder architecture. This architecture has been widely explored in speech recognition and various NLP tasks [31][32] but to the best of our knowledge has not been applied to depression prediction problems. We refer to this model as the EH-AC model. The EH-AC model consists of multiple layers of CNNs and LSTMs (collectively the "encoder") and a prediction network (classifier or regression) over it. Different architectures can be used for the prediction network; here we present only the results of a Recurrent CNN [33] (RCNN) model.

The EH-AC model is fed with filter-bank coefficients which are computed with an analysis window of 25 ms and a frame rate of 10 ms. After experimentation, we found that these values work relatively well across different speech and audio applications, though further tuning could be done in future work. Due to the memory limitations of our GPU devices, it was not possible to include the full duration of a speaker's session at once. Instead, we created shorter time segments from each session. We found that longer segments perform better than shorter segments, but longer segments also require more memory. After experimenting with this trade-off, we found that 25 seconds provided the best performance on our development set while staying within the memory constraints for our hardware. Longer durations could be explored given higher capacity hardware resources.

An overview of the EH-AC model is provided in Figure 1 (solid arrows). The speech signal is first divided into segments with a maximum length of 25 seconds. These segments are then passed to the feature extraction module described above. The resulting features then pass through the encoder and deep prediction network to obtain a prediction for each individual segment. To compute a session-level prediction, segment-level outputs are aggregated using a segment fusion module, which is discussed in section 2.2.3.

#### 2.2.2. Transfer learning and comparison to prior work

Transfer learning is applied to the EH-AC model by adding a decoder module to the network. The most similar other work that we have found that applies transfer learning and deep architecture to the task of predicting depression from speech signal is reported in [12]. Our model is different from [12] in its overall architecture, number of modules (our model has less modules), training steps (our model has fewer steps), and the transfer learning approach. In [12], the authors applied both unsupervised and supervised source tasks in their pretraining stage and then transfer the attention weights. In contrast, we apply only a supervised task (ASR) and then transfer only the weights of the encoder.

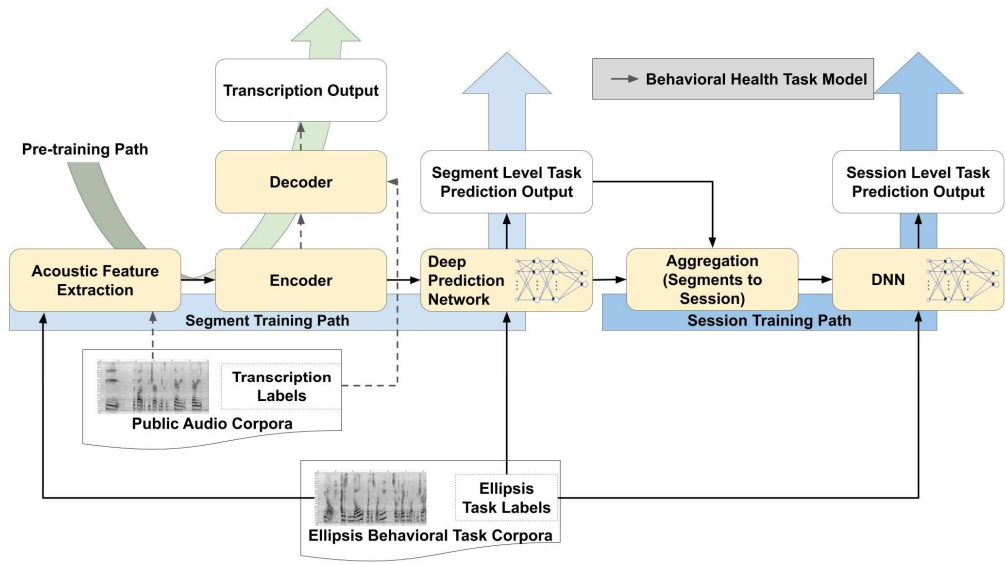

**Figure 1** An overview of the model. Transfer learning is achieved using an ASR task (dashed arrows). Depression prediction is performed using a pretrained encoder network (solid arrows).

In Figure 1, the transfer learning path is shown with dashed arrows to indicate that it only applies to the pretraining stage. The transfer learning path is removed once the encoder is pretrained. The decoder consists of an LSTM layer with attention. Automatic Speech Recognition (ASR) is used as the source task. In the pretraining stage, we train the encoder and decoder with transcribed speech data (unlabeled for behavioral health). After pretraining, we remove the decoder (dashed arrows) and train the rest of the network as mentioned in the previous section using labeled data. The ASR decoder is computationally more expensive than the encoder and prediction modules. Our model is relatively lightweight during inference since we can remove the ASR decoder. This is an important practical advantage of our model during deployment.

The ASR sub-model is based on a hybrid connectionist temporal classification (CTC)/attention architecture [31] and is inspired by prior work including that in [31], [34], and [35]. To train the ASR task, we used the Librispeech dataset [36], which is an English dataset comprising over 1000 hours of read speech. We note that speaking style is unmatched with respect to our data which comprise spontaneous speech samples in which users talk freely about their lives. Thus, we did not expect the model pretrained on Librispeech to perform well on our data. Our goal, rather, was for our model to learn a representation for the acoustic space. We plan to investigate corpora in addition to Librispeech in future work; this will reveal whether style match is important for the impact of transfer learning in our domain.

Past work has shown that transfer learning can improve the performance of machine learning algorithms on new tasks by leveraging data and feature representations learned from other well-studied tasks [23][37]. We assume that, by pretraining our encoder, our network is forced to learn a more restrictive representation relative to when we trained all layers from scratch; that is, we assume that the first few layers act as an advanced feature extractor for the predictor.

### 2.2.3. Segment fusion

We fuse the individual segments described in 2.2.1 using an additional neural network. Every segment is represented by a vector corresponding to the last hidden layer of the prediction subnetwork (e.g. RCNN in this case). The sequence of segments for every session is projected into a single vector by max pooling and is then fed into an MLP network. The model then can be trained for either classification or regression tasks and the output is interpreted accordingly. The output of this sub-module is a prediction for the overall session.

## 3. RESULTS & DISCUSSION

In this section we present results for both classification and regression tasks. We report results on both development and test splits. Model parameters were selected using only the development set; the test set was used only for the final evaluation. The final prediction for each session was computed using the segment fusion module and evaluated using the session PHQ-8 label either as a binary class (+dep versus -dep) for classification or directly as a PHQ-8 value for regression.

Table 2 Binary classification results for depression prediction.
Models with transfer learning are indicated using *+TL*.

| Model | AUC | Specificity at EER | Sensitivity at EER |
|---|---|---|---|
| EH-AC+TL-1/dev | 0.78 | 0.70 | 0.70 |
| EH-AC+TL-1/test | 0.79 | 0.71 | 0.71 |
| EH-AC+TL-2/dev | 0.77 | 0.71 | 0.71 |
| EH-AC+TL-2/test | 0.79 | 0.72 | 0.72 |
| EH-AC/dev | 0.62 | 0.58 | 0.58 |
| EH-AC/test | 0.63 | 0.59 | 0.59 |
| CNN/dev | 0.60 | 0.59 | 0.59 |
| CNN/test | 0.60 | 0.56 | 0.56 |
| LSTM/dev | 0.61 | 0.59 | 0.59 |
| LSTM/test | 0.58 | 0.56 | 0.56 |

Table 2 shows the results of the binary classification task for several models. Specificity and sensitivity are calculated at the equal error rate (EER) point. Area under the curve (AUC) is reported as the single metric to compare different models. The statistical significance of differences in AUC is calculated using the DeLong test [38]. We also show the results of an EH-AC model trained from scratch. Results using CNN (with six convolutional layers and two fully connected layers) and LSTM (with two LSTM and two fully connected layers) models are also included. CNN and LSTM models are among the most-used models and have been applied in earlier studies of depression prediction, including [8][10][29][39]. All the models listed perform significantly better than chance (in a DeLong test at $p < 0.05$). CNN, LSTM, and EH-AC models all achieve AUC close to 0.60 and a DeLong test shows that the difference between models is not statistically significant. The EH-AC with transfer learning (EH-AC+TL) model, however, provides a relative 27% gain over the EH-AC baseline. The DeLong test also shows that adding transfer learning to the EH-AC model results in highly significant improvement with a p-value close to zero.

To better understand the effect of transfer learning, we designed two experiments. In the first experiment (denoted by EH-AC+TL-1) we trained an ASR task in the pretraining stage by updating both encoder and decoder weights. This resulted in a relatively "strong" source task. In the second experiment (denoted by EH-AC+TL-2), we did not update the decoder weights and the result was a "weak" source task. The character error rate (CER) is 30% for experiment 1 and 188% for experiment 2 (due to insertion errors, CER can exceed 100%). Both ASR tasks can be considered weak relative to state-of-the-art ASR models, but clearly one is much weaker than the other. We can see that the gain from transfer learning for these models is virtually the same (not statistically significant under DeLong test); that is, interestingly, most of the gain in performance can be achieved with even a weak ASR task for the pretraining step.

Table 3 shows the performance for the corresponding regression task, i.e. directly predicting PHQ-8 results, without class mapping. Results are shown for root mean square error (RMSE), mean square error (MAE), and Pearson correlation (PCC). LSTM and CNN results are not shown but, as was the case with binary classification, results are similar to those for the EH-AC model. Transfer learning also results in an improvement for regression performance, with a 11% relative reduction in RMSE and 13% relative reduction in MAE. Both binary classification and regression results show remarkably stable performance over development and test sets. This is in contrast to past results using smaller datasets [8][12] in which performance changes widely from development to test set.

Table 3 Regression results for the depression prediction problem.

| Model | RMSE | MAE | PCC |
|---|---|---|---|
| EH-AC+TL/dev | 4.60 | 3.47 | 0.51 |
| EH-AC+TL/test | 4.70 | 3.56 | 0.49 |
| EH-AC/dev | 5.25 | 4.11 | 0.18 |
| EH-AC/test | 5.26 | 4.12 | 0.21 |

### 4. CONCLUSIONS & FUTURE WORK

Using a large labeled corpus, we investigated a lightweight encoder and encoder-weight-only transfer learning approach for the task of predicting depression using acoustic information. The transfer learning is based on an ASR task. The run-time model is lightweight because the ASR decoder is removed after the pretraining stage.

Transfer learning based on this method results in a 27% relative performance boost for binary classification and a 10-15% relative reduction in regression error metrics. By using a large dataset, we obtain robust results with no difference between test and development set performance. This robustness is partially the result of applying a large dataset and of transfer learning. However, it is also a result of the matched data distribution. In the future, we plan to extend this study by evaluating our models on new test data with more variety relative to our current dataset.

We also found that a weakly-performing ASR source task in the transfer learning phase added almost the same gain in performance as a better-performing ASR task. Taken together, these results suggest that the approach is flexible and offers promise for efficient implementation. We plan to study the effect of source task performance, source task type (e.g. unsupervised tasks), and source task data in future work.

### 5. ACKNOWLEDGMENTS

We thank Mainul Mondal, Mike Aratow, David Lin and Tahmida Nazreen for support and contributions.

### 6. REFERENCES

[1] World Health Organization, "Depression and other Common Mental Disorders: Global Health Estimates," World Health Organization, pp. 1–24, 2017.

[2] American Psychiatric Association, "Major depressive disorder," in Diagnostic and Statistical Manual of Mental Disorders, 5th ed., Arlington, VA, USA: American Psychiatric Association, 2013.

[3] Covid1 World Health Organization, "Coronavirus disease 2019 (COVID-19): Situation report 82," WHO, 2020.

[4] Y. Huang and N. Zhao, "Generalized anxiety disorder, depressive symptoms and sleep quality during COVID-19 outbreak in China: a web-based cross-sectional survey," Psychiatry Res., p. 112954, 2020.

[5] J. R. Williamson et al., "Detecting Depression using Vocal, Facial and Semantic Communication Cues," in Proc. 6th Int. Workshop Audio/Visual Emotion Challenge, 2016, pp. 11–18.

[6] P. Resnik, A. Garron, and R. Resnik, "Using Topic Modeling to Improve Prediction of Neuroticism and Depression," in Proc. 2013 Conf. Empirical Methods Nat. Lang. Process., 2013, pp.1348–1353.

[7] A. Pampouchidou et al., "Depression Assessment by Fusing High- and Low-Level Features from Audio, Video, and Text," in Proc. 6th Int. Workshop Audio/Visual Emotion Challenge, 2016, pp. 27–34.

[8] F. Ringeval et al. "AVEC 2019 Workshop and Challenge: State-of-mind, Detecting Depression with AI, and Cross-Cultural Affect Recognition," in Proc. 9th Int. Audio/Visual Emotion Challenge and Workshop, 2019, pp. 3–12.

[9] N. Cummins, S. Scherer, J. Krajewski, S. Schnieder, J. Epps, and T. F. Quatieri. "A Review of Depression and Suicide Risk Assessment Using Speech Analysis," Speech Commun., vol. 71, pp. 10–49, 2015.

[10] L. Yang, H. Sahli, X. Xia, E. Pei, M. C. Oveneke, and D. Jiang. "Hybrid Depression Classification and Estimation from Audio, Video and Text Information," in Proc. 7th Annu. Workshop Audio/Visual Emotion Challenge, 2017, pp. 45–51.

[11] T. Rutowski, A. Harati, Y. Lu, and E. Shriberg, "Optimizing Speech-Input Length for Speaker-Independent Depression Classification," in Proc. Interspeech, 2019, pp. 3023–3027.

[12] Z. Zhao, Z. Bao, Z. Zhang, N. Cummins, H. Wang, and B Schuller. "Hierarchical Attention Transfer Networks for Depression Assessment from Speech," in IEEE Int. Conf. Acoust., Speech and Signal Process., 2020, pp. 7159–7163.

[13] G. Coppersmith, "CLPsych 2015 Shared Task: Depression and PTSD on Twitter," in Proc. 2nd Workshop Comput. Ling. Clin. Psychol., 2015, pp. 31–39.

[14] M. Valstar, et al., "AVEC 2016: Depression, Mood, and Emotion Recognition Workshop and Challenge," in Proc. 6th Int. Workshop Audio/Visual Emotion Challenge, 2016, pp. 3–10.

[15] F. Ringeval, et al., "AVEC 2017: Real-life Depression and Affect Recognition Workshop and Challenge," in Proc. 7th Annu. Workshop Audio/Visual Emotion Challenge, 2017, pp. 3–9.

[16] J. F. Cohn, N. Cummins, J. Epps, R. Goecke, J. Joshi, and S. Scherer. "Multimodal Assessment of Depression from Behavioral Signals," in The Handbook of Multimodal-Multisensor Interfaces: Signal Processing, Architectures, and Detection of Emotion and Cognition, vol. 2, pp. 375–417, 2018.

[17] S. Scherer, G. Stratou, J. Gratch, and L-P. Morency, "Investigating Voice Quality as a Speaker-Independent Indicator of Depression and PTSD," in Proc. Interspeech, 2013, pp. 847–851

[18] B. Stasak, J. Epps, and R. Goecke. "An Investigation of Linguistic Stress and Articulatory Vowel Characteristics for Automatic Depression Classification." Comput. Speech Lang., pp. 140–155, 2019.

[19] A. Trevino, T. Quatieri, and N. Malyska, "Phonologically-Based Biomarkers for Major Depressive Disorder," EURASIP J. Adv. Signal Process., pp. 1–18, 2011.

[20] B. Helfer, T. Quatieri, J. R. Williamson, D. Mehta, R. Horwitz, and B. Yu, "Classification of Depression State Based on Articulatory Precision," in Proc. Interspeech, 2013, pp. 2172–2176.

[21] I. Goodfellow, Y. Bengio, and A. Courville, Deep Learning. MIT Press, 2016.

[22] Q. Mao, M. Dong, Z. Huang, and Y. Zhan, "Learning Salient Features for Speech Emotion Recognition using Convolutional Neural Networks." IEEE Trans. Multimedia, vol. 16, no. 8, pp. 2203–2213, 2014.

[23] J. Deng, Z. Zhang, E. Marchi, and B. Schuller, "Sparse Autoencoder-Based Feature Transfer Learning for Speech Emotion Recognition," in Humaine Assoc. Conf. Affect. Comput. and Intell. Interact., 2013, pp. 511–516.

[24] W. Lim, D. Jang, and T. Lee, "Speech Emotion Recognition Using Convolutional and Recurrent Neural Networks," in Asia-Pacific Signal Inf. Process. Assoc. Annu. Summit and Conf., 2016, pp. 1–4.

[25] S. E. Kahou et al., "Emonets: Multimodal Deep Learning Approaches for Emotion Recognition in Video," J. Multimodal User Interfaces, vol. 10, no. 2, pp. 99–111, 2016.

[26] Z. Huang, M. Dong, Q. Mao, and Y. Zhan, "Speech Emotion Recognition using CNN," in Proc. 22nd ACM Int. Conf. Multimedia, 2014, pp. 801–804.

[27] L. Yang, D. Jiang, and H. Sahli. "Feature Augmenting Networks for Improving Depression Severity Estimation from Speech Signals." IEEE Access 8, pp. 24033–24045, 2020.

[28] Z. Huang, J. Epps, and D. Joachim. "Exploiting Vocal Tract Coordination using Dilated CNNS for Depression Detection in Naturalistic Environments," in IEEE Int. Conf. Acoust., Speech and Signal Process., 2020, pp. 6549–6553.

[29] L. He and C. Cao, "Automated Depression Analysis using Convolutional Neural Networks from Speech," J. Biomed. Inf., vol. 83, pp. 103–111, 2018.

[30] K. Kroenke, T.W. Strine, R. Spitzer, J.B.W. Williams, J.T. Berry, and A.H. Mokdad, "The PHQ-8 as a Measure of Current Depression in the General Population," J. Affect. Disord., vol. 114, no. 1–3, 2009.

[31] S. Kim, T. Hori, and S. Watanabe, "Joint CTC-Attention based End-to-End Speech Recognition using Multi-Task Learning," in IEEE Int. Conf. Acoust., Speech and Signal Process., 2017, pp. 4835–4839.

[32] I. Sutskever, O. Vinyals, and Q. V. Le. "Sequence to sequence learning with neural networks," Adv. Neural Inf. Process. Syst., pp. 3104–3112, 2014.

[33] S. Lai, L. Xu, K. Liu, and J. Zhao. "Recurrent Convolutional Neural Networks for Text Classification," in 29th AAAI Conf. Artif. Intell., 2015, pp. 2267–2273.

[34] W. Chan, N. Jaitly, Q. Le, and O. Vinyals, "Listen, Attend and Spell: A Neural Network for Large Vocabulary Conversational Speech Recognition," in IEEE Int. Conf. Acoust., Speech and Signal Process., 2016, pp. 4960–4964.

[35] A. H. Liu, T-W. Sung, S-P. Chuang, H-Y. Lee, and L-S. Lee, "Sequence-to-sequence Automatic Speech Recognition with Word Embedding Regularization and Fused Decoding," in IEEE Int. Conf. Acoust., Speech and Signal Process., 2020, pp. 7879–7883.

[36] V. Panayotov, G. Chen, D. Povey, and S. Khudanpur. "Librispeech: an ASR Corpus Based on Public Domain Audio Books," in IEEE Int. Conf. Acoust., Speech and Signal Process., 2015, pp. 5206–5210.

[37] X. Glorot, A. Bordes, and Y. Bengio, "Domain Adaptation for Large-Scale Sentiment Classification: A Deep Learning Approach," ICML, 2011, pp. 513–520.

[38] E. DeLong, D. DeLong, and D. Clarke-Pearson. "Comparing the areas under two or more correlated receiver operating characteristic curves: a nonparametric approach," Biometrics, pp. 837–845, 1988.

[39] T. Alhanai, M. Ghassemi, and J. Glass, "Detecting Depression with Audio/Text Sequence Modeling of Interviews," in Proc. Interspeech, 2018, pp. 1716–1720.